\documentclass[conference]{IEEEtran}
\IEEEoverridecommandlockouts

\usepackage{cite}
\usepackage{amsmath,amssymb,amsfonts}
\usepackage{algorithmic}
\usepackage{graphicx}
\usepackage{textcomp}
\usepackage{xcolor}
\usepackage{tabularx}
\usepackage{stfloats}
\def\BibTeX{{\rm B\kern-.05em{\sc i\kern-.025em b}\kern-.08em
    T\kern-.1667em\lower.7ex\hbox{E}\kern-.125emX}}
\renewcommand{\baselinestretch}{0.97}
\begin{document}
\title{Localized Thermal Management of Induction-Heated Microrobots for Hyperthermia Applications\thanks{\textsuperscript{1}A. Balaji is an NSF REM Scholar at the University of Houston and a student at West Windsor-Plainsboro High School North, NJ, USA. \textsuperscript{2}F. Arevalo and A. T. Becker are with the University of Houston.}}
\author{
    \IEEEauthorblockN{Advika Balaji\textsuperscript{1}, Miguel F. Arevalo-Castiblanco\textsuperscript{2}, and Aaron T. Becker\textsuperscript{2}}
    \IEEEauthorblockA{\textit{Department of Electrical and Computer Engineering, University of Houston, Houston, TX, USA} \\
    Emails: abalaji3@uh.edu, fareval2@uh.edu, atbecker@uh.edu}
}

\maketitle
\begin{abstract}
Induction-heated microrobots face a fundamental thermal management challenge: the alternating magnetic fields required for localized hyperthermia simultaneously generate parasitic ohmic heating within the actuation coils. Conventional cooling approaches can fail because conductive cooling structures placed near the coil can couple to the time-varying field, generating additional electromagnetic losses. This paper addresses this architectural limitation by evaluating two liquid-cooling frameworks designed to isolate the electromagnetic workspace: an indirect configuration utilizing an external copper water block, and a direct architecture routing coolant internally through a hollow helical coil. Evaluated over a 100~s window with multi-point thermocouple telemetry, the indirect water-block configuration failed due to secondary inductive coupling with the block itself, acting as an unintended susceptor that drove the block temperature to an extreme peak of 805$^{\circ}$C. In contrast, the direct internal cooling architecture successfully extracted ohmic losses at the source, sustaining a localized ablation tip temperature of 63.4$^{\circ}$C while strictly restricting surrounding fluid and boundary surfaces below defined thresholds of 33.9$^{\circ}$C and 37.6$^{\circ}$C, respectively. These results establish coil-level fluidic integration as a key design principle for thermally isolated microrobot actuation and provide a hardware foundation for future spatially controlled hyperthermia.
\end{abstract}


\section{Introduction}
The development of minimally invasive medical technologies has increasingly focused on reducing the physical footprint of therapeutic devices while improving their ability to reach and treat specific regions of the body. Conventional medical instruments are often limited by their size, geometry, and reliance on direct access to a treatment site. These limitations become especially important when the target is located within a narrow, difficult-to-reach anatomical environment.

Miniature Magnetic Robotic Swimmers (MMRS) offer a possible solution. Because their movement can be controlled using externally applied magnetic fields, they do not necessarily require a physical connection between the device and an external actuator. Prior studies have explored MMRS during targeted delivery, cell manipulation, imaging, and minimally invasive surgery ~\cite{nelson2010microrobots}. Their compactness makes them particularly attractive for procedures where conventional tools are difficult to maneuver.
\begin{figure}[!b]
  \centering
  \includegraphics[width=\columnwidth]{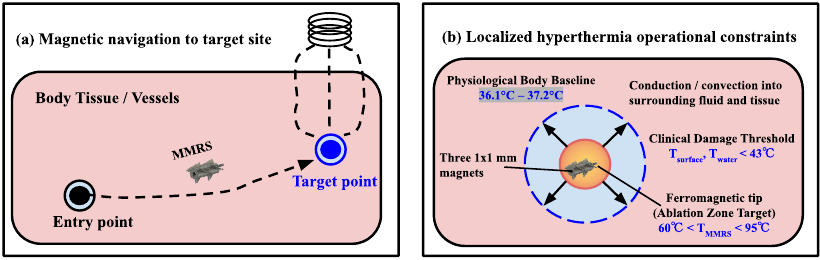}
  \caption{Concept overview of the untethered microrobotic platform: (a) electromagnetic navigation of the MMRS to the target site, and (b) localized induction thermal ablation operational constraints relative to the human physiological baseline.}
  \label{fig:concept}
\end{figure}
One potential application of this technology is localized thermal therapy, notably hyperthermia. This procedure is the clinical elevation of tissue temperature above normal levels to damage or destroy abnormal tissue. Magnetic hyperthermia commonly uses magnetic nanoparticles exposed to an alternating magnetic field, converting electromagnetic energy into heat. This technique has been assessed during cancer therapy, drug delivery, and other biomedical applications where heating is required ~\cite{liu2020comprehensive}. 

Unlike nanoparticle-based magnetic hyperthermia, which distributes heat-generating particles throughout a treatment region, an induction-powered MMRS acts as a localized heat source. This architecture combines magnetic navigation and localized heating within a single untethered device, as illustrated in Fig.~\ref{fig:concept}.

Although MMRS research has advanced rapidly in biomedical applications, existing work has primarily focused on propulsion, navigation, drug delivery, imaging, and multifunctional structural designs~\cite{sitti2015untethered}. In parallel, magnetic hyperthermia research has focused on magnetic materials, heat generation, and treatment efficacy~\cite{liu2020comprehensive}. The intersection of these areas introduces a distinct engineering challenge: unwanted collateral heating within the actuation workspace.

This is particularly relevant for thermal ablation, where temperatures required for treatment can exceed the tolerance of surrounding biological tissue. Heat generated at the swimmer tip can propagate through the surrounding fluid and boundaries by conduction and convection, producing unintended thermal exposure outside the target region. Consequently, system performance depends not only on the maximum achievable tip temperature, but also on the ability to control the surrounding thermal field.

To address this challenge, we experimentally evaluate two active cooling architectures for an induction-heated MMRS platform. The main contributions of this work are:
\begin{itemize}
    \item \textbf{Thermal Characterization:} We show external conductive cooling elements act as an unintended susceptor, causing substantial parasitic heating within the actuation workspace.
    \item \textbf{Field-Preserving Fluidics:} We implement a hollow helical coil with internal coolant flow to remove parasitic ohmic heat at the source while preserving the external magnetic actuation field.
    \item \textbf{Robotic Hardware Baseline:} We demonstrate localized tip heating up to 63.4$^{\circ}$C while maintaining surrounding fluid and boundaries below critical clinical thresholds at 33.9$^{\circ}$C and 37.6$^{\circ}$C, respectively, establishing a thermal baseline for future spatially regulated therapy.
\end{itemize}
The remainder of this paper is organized as follows: Section~II reviews the technical background on MMRS biomedical applications and magnetic heating. Section~III outlines the experimental materials, methods, and the dual liquid-cooling configurations. Section~IV presents the empirical thermal results. Section~V discusses experimental implications and system limitations. Finally, Section~VI concludes the paper.

\section{Background}

\subsection{MMRS for Biomedical Applications}
Magnetic actuation is particularly attractive for medical microrobotics because the driving field can be applied externally. Unlike chemically powered micromotors, magnetic systems do not require the robot to carry a conventional fuel source~\cite{sanchez2015chemically}. Magnetic control can also provide multiple forms of motion. Depending on the geometry of the robot and the applied magnetic field, magnetic forces and torques can be used to orient and translate the device. For example, helical microrobots can convert rotational magnetic actuation into translational motion through a corkscrew-like swimming motion~\cite{abbott2009how,ghosh2009controlled}. The broader field includes helical, surface-based, ciliary, scaffold, and biohybrid designs~\cite{sitti2015untethered,ricotti2017biohybrid}.

For biomedical use, however, locomotion is only one part of the problem. A clinically useful microrobot must also address biocompatibility, tracking, control, reliability, fabrication, and interaction with biological tissue. Past papers have emphasized that translating MMRS from laboratory evaluation to clinical systems requires integration of mechanical, magnetic, imaging, and control systems~\cite{nelson2010microrobots}. Our systems perspective is particularly relevant to thermal microrobotics. A swimmer that can navigate accurately but not control its temperature would not be sufficient for thermal therapy.

\subsection{Hyperthermia and Ablation Trajectories}
Magnetic hyperthermia clinically elevates tissue temperatures to induce targeted thermal necrosis~\cite{liu2020comprehensive}. While conventional approaches distribute functional nanoparticles throughout a target volume~\cite{liu2020comprehensive}, this untethered robotic system utilizes a dual-mechanism architecture—combining remote magnetic manipulation via embedded permanent magnets with localized inductive heating at the ferromagnetic iron tip—to function as a steerable, heat-generating ablation source. Standard procedures like Radiofrequency Ablation (RFA) effectively induce cell necrosis above 50$^{\circ}$C but mandate invasive physical catheter path access, introducing vascular risks and uncontrolled outward thermal conduction that endangers adjacent sensitive anatomical structures~\cite{goldberg2001radiofrequency,ahmed2011principles}. Remote, untethered microrobotic architectures effectively mitigate these mechanical access risks, provided that localized workspace heat dissipation can be strictly bounded to prevent collateral damage to healthy peripheral tissue.

\section{Materials and Methods}
\begin{table}[!t]
\caption{System Specifications and Geometric Parameters}
\label{tab:system_specs}
\centering
\footnotesize
\setlength{\tabcolsep}{4pt}
\renewcommand{\arraystretch}{1.2}
\begin{tabular}{|>{\raggedright\arraybackslash}p{2.1cm}|>{\raggedright\arraybackslash}p{2.4cm}|>{\raggedright\arraybackslash}p{2.7cm}|}
\hline
\textbf{Category} & \textbf{Component / Property} & \textbf{Value / Specification} \\
\hline
\textbf{Magnetic Actuation} & Induction Power Source & Hot Rod H7E (1000~W) \\ \cline{2-3}
& Coil Tube Outer OD & 5.0 mm \\ \cline{2-3}
& Coil Turns & 6.5 turns \\ \hline
\textbf{MMRS Architecture} & Swimmer Length & 30.0 mm \\ \cline{2-3}
& Tip Inner Radius & 8.0 mm \\ \cline{2-3}
& Tip Outer Radius & 9.0 mm \\ \cline{2-3}
& Permanent Magnets & 3 units ($1 \times 1$~mm) \\ \hline
\textbf{Fluidic Cooling} & Coolant Fluid & Distilled Water (35 g) \\ \cline{2-3}
& Pump Flow Rate & 300 L/h (12 V, 4 W) \\ \cline{2-3}
& Radiator Assembly & 12~V, 0.15~A Fan \\ \hline
\textbf{Data Acquisition} & Temperature Sensors & Type E Thermocouples \\ \cline{2-3}
& Hardware Controller & NI cRIO Modules \\ \cline{2-3}
& Software Interface & LabVIEW Baseline \\ \hline
\end{tabular}
\end{table}
\subsection{Experimental System}
The experimental system was developed to regulate heat generated during induction-based MMRS operations utilizing a closed-loop fluidic architecture inspired by liquid-cooled computing hardware. The thermal energy generated by localized high-power components is transferred into a circulating liquid coolant, transported away from the heat source, and dissipated into the ambient environment via a radiator and fan assembly. In both configurations, the loop consists of a water pump, flexible silicone tubing, a primary heat-transfer interface, and a heat-rejection radiator.
\begin{figure}[!t]
  \centering
  \includegraphics[width=0.95\columnwidth]{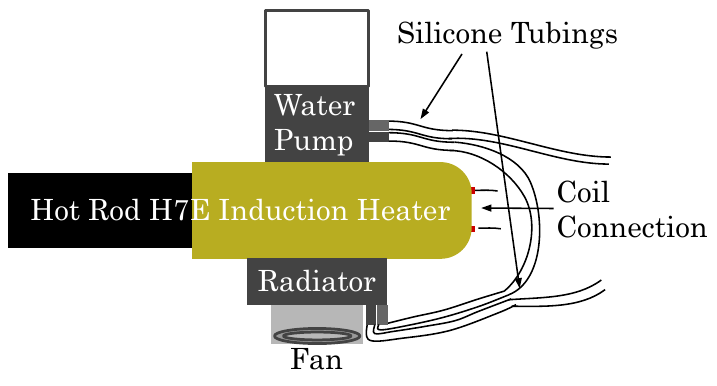}
  \caption{\fontsize{8}{10}\selectfont Schematic of the experimental system infrastructure and circulating fluidic loop layout.}
  \label{fig:concept2}
\end{figure}
The structural configuration varies by the location and mechanism of heat extraction from the induction process, as illustrated in Fig.~\ref{fig:concept2}. The comprehensive hardware specifications and geometric parameters for both configurations are detailed in Table~\ref{tab:system_specs}. The architecture operates on the premise that the induction coil functions as both the primary electromagnetic power source and a core node of parasitic thermal losses. Consequently, active thermal management must target the coil structure directly, extracting waste heat at the source to prevent conduction and convection from elevating surrounding temperatures.

\subsection{MMRS Architecture}
\begin{figure*}[!t]
  \centering
  \begin{minipage}{0.48\linewidth}
    \centering
    \includegraphics[width=0.60\linewidth]{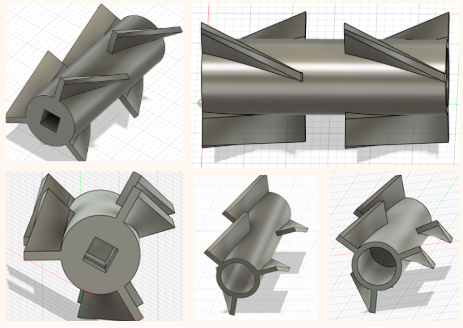}
    \caption{Structural architecture of the SLA-printed MMRS defining the dual-functional propeller fins.}
    \label{fig:mmrs_architecture}
  \end{minipage}\hfill
  \begin{minipage}{0.48\linewidth}
    \centering
    \includegraphics[width=1.0\linewidth]{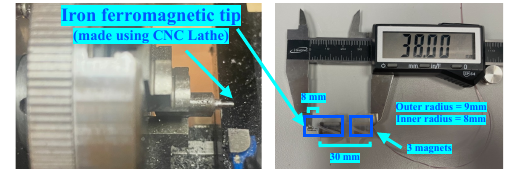}
    \caption{Fabricated MMRS housing showcasing the lathe-turned iron ferromagnetic tip and longitudinal geometric dimensions.}
    \label{fig:mmrs_housing}
  \end{minipage}
\end{figure*}
The posterior cavity houses three 1$\times$1~mm permanent magnets bonded with a high-strength two-part epoxy. The body features four angled propeller fins that convert the rotational torque from an externally applied rotating magnetic field into axial thrust about the longitudinal axis. This unified magnetic actuation framework enables the MMRS to seamlessly transition between orientation control and translational locomotion through confined pathways. Additionally, these integrated fins provide an expanded surface area to enhance fluidic convective heat transfer away from the robot.

\subsection{Induction Heating Subsystem}
A commercial Hot Rod H7E 1,000~W induction heater was used to generate an alternating electromagnetic field, driving current through the copper induction coil to produce a time-varying magnetic field in the region occupied by the MMRS. The amount of energy transferred to the MMRS depends on the electromagnetic coupling between the coil and the ferromagnetic tip, which is governed by the coil geometry, separation distance, applied excitation voltage, and material properties. The induction coil itself also undergoes resistive heating because of the high current required to generate the alternating magnetic field, where its thermal power can be approximated by \cite{rudnev2017handbook}:
\begin{equation} 
P_{\mathrm{coil}} = I_{\mathrm{rms}}^{2} R_{\mathrm{AC}}(f), 
\label{eq:coil_power} 
\end{equation}
where $I_{\mathrm{rms}}$ represents the root-mean-square current supplied to the induction coil, and $R_{\mathrm{AC}}(f)$ denotes the AC resistance as a function of the operating frequency $f$. For the physical characterizations presented herein, the magnetic drive system was energized via the 1,000~W induction power source operating at its resonant tank frequency, depositing high-frequency current through the copper geometry to establish the time-varying field.

Therefore, the induction subsystem produces localized heating at the MMRS tip while simultaneously generating parasitic thermal losses within the induction coil. This dual thermal response creates a fundamental constraint in the system design: increasing electromagnetic excitation increases power deposited in the MMRS and its achievable temperature, but also escalates the electrical losses and resulting thermal load within the coil. The power distribution can be represented conceptually as:
\begin{equation}
P_{\mathrm{ind}} = P_{\mathrm{MMRS}} + P_{\mathrm{coil}} + P_{\mathrm{loss}},
\label{eq:power_distribution}
\end{equation}
where $P_{\mathrm{ind}}$ is the electrical power supplied to the induction system, $P_{\mathrm{MMRS}}$ represents the power deposited in the ferromagnetic MMRS tip through electromagnetic losses, $P_{\mathrm{coil}}$ denotes resistive losses within the copper coil, and $P_{\mathrm{loss}}$ accounts for energy dissipated to the surrounding environment. The cooling architectures were therefore designed to extract heat generated by the induction coil and prevent thermal deformation while maintaining the electromagnetic coupling necessary for effective MMRS heating.

\subsection{Liquid Cooling Architecture}
The thermal management loop adapts commercial-off-the-shelf components to regulate the induction system's high heat flux, preventing thermal deformation to enable continuous hardware operation. Similar to electronics cooling, a circulating fluid loop transfers heat from the induction coil into a sealed circuit consisting of a 150~mL, 12~V, 4~W, 300~L/h water pump, flexible silicone tubing secured with iron hose clamps, and an 80~mm, 12~V, 0.15~A radiator and fan assembly. Distilled water serves as the coolant due to its high specific heat capacity of 4.184~J/(g$\cdot^{\circ}$C), absorbing large thermal loads with minimal fluctuations. The thermal energy transport rate away from the induction workspace is modeled as~\cite{bergman2017fundamentals}:
\begin{equation}
Q_{\mathrm{c}} = \dot{m}c_{\mathrm{p}}(T_{\mathrm{out}} - T_{\mathrm{in}}),
\label{eq:coolant_heat_transfer}
\end{equation}
where $\dot{m} = \rho \dot{V}$ is the coolant mass-flow rate—derived from the volumetric flow rate $\dot{V}$ listed in Table~\ref{tab:system_specs} and the fluid density $\rho$—$c_{\mathrm{p}}$ is the specific heat capacity, and $T_{\mathrm{in}}$ and $T_{\mathrm{out}}$ are the fluid temperatures entering and exiting the heat-transfer region, respectively. By driving continuous volumetric flow through the tubing and radiator, the pump prevents localized heat accumulation and reduces the risk of thermal degradation of the induction coil components.

Flexible silicone tubing, secured with iron hose clamps, provides the fluidic routing between the pump, heat exchanger, and radiator. The radiator serves as the primary heat-rejection component, where thermal energy conducts through the internal walls into an external fin structure. An 80~mm fan drives forced convection across these fins to accelerate dissipation into the environment at a rate governed by:
\begin{equation}
Q_{\mathrm{conv}} = hA(T_{\mathrm{s}} - T_{\infty}),
\label{eq:convective_heat_transfer}
\end{equation}
where $h$ is the convective heat-transfer coefficient, $A$ is the effective heat-transfer area, $T_{\mathrm{s}}$ is the radiator surface temperature, and $T_{\infty}$ is the ambient air temperature.

\subsection{Configuration 1: Water Block Cooled}
\begin{figure*}[!t]
  \centering
  \begin{minipage}{0.48\linewidth}
    \centering
    \includegraphics[width=0.85\linewidth]{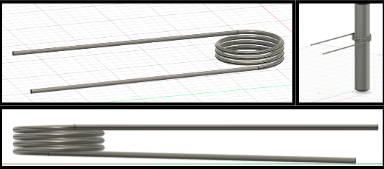}
    \caption{Structural layout of the solid copper helical coil in Configuration 1.}
    \label{fig:water_block_coil}
  \end{minipage}\hfill
  \begin{minipage}{0.48\linewidth}
    \centering
    \includegraphics[width=0.95\linewidth]{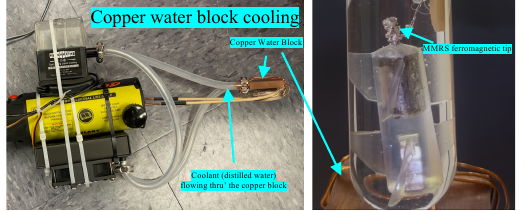}
    \caption{Physical experimental configuration of the external copper water-block loop assembly.}
    \label{fig:water_block_assembly}
  \end{minipage}
  \vspace{-1.0em} 
\end{figure*}

The first experimental configuration used a solid copper helical coil with an inner diameter of 20~mm, wire thickness of 2~mm, outer diameter of 24~mm, and a mean coil diameter of 22~mm, yielding a total wire length of 345~mm, as shown in Fig.~\ref{fig:water_block_coil}.
\begin{figure*}[!b]
  \centering
  \begin{minipage}{0.48\linewidth}
    \centering
    \includegraphics[width=0.97\linewidth]{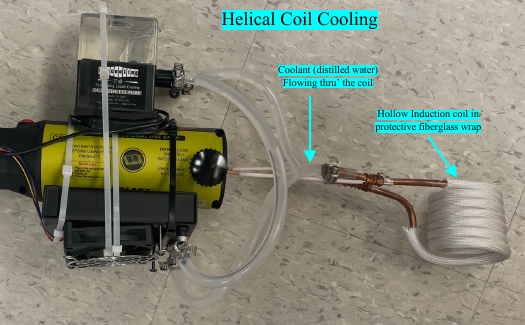}
    \caption{Direct internal water-cooling routed seamlessly through the hollow helical induction coil.}
    \label{fig:direct_water_cooling}
  \end{minipage}\hfill
  \begin{minipage}{0.48\linewidth}
    \centering
    \includegraphics[width=0.80\linewidth]{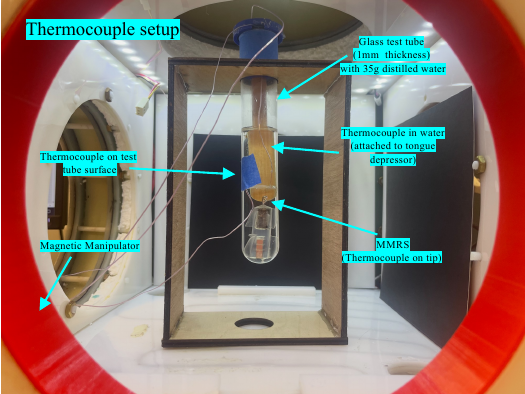}
    \caption{Spatial positioning and node distribution of the multi-point thermocouple telemetry.}
    \label{fig:thermocouple_telemetry}
  \end{minipage}
\end{figure*}
In this configuration, heat generated by the coil conducts across the physical contact interface into the copper water block, where it is subsequently removed via fluid convection. The thermal dissipation efficiency is inherently constrained by the contact surface area and interfacial thermal resistance. This approach mirrors localized cooling methodologies common in conventional electronics packaging.

\subsection{Configuration 2: Directly Water Cooled \& Instrumentation Layout}
The second configuration minimizes the thermal path by routing the coolant directly through a hollow helical induction coil, eliminating the secondary water-block interface, as shown in Fig.~\ref{fig:direct_water_cooling}.
This hollow coil featured a 56~mm mean diameter (50~mm inner diameter, 62~mm outer diameter), 6.5~turns (a 42~mm height), using a tube with a 6~mm outer diameter and a 2.5~mm internal channel diameter. Tubing was connected directly to the coil inlet and outlet. To prevent ambient heat radiation and ensure operator safety, the hollow coil was encased in a protective fiberglass wrap. Coolant enters the conductor channel, absorbs heat directly from the internal walls of the energized copper coil, and carries the thermal energy to the radiator, bypassing interfacial losses.

\subsection{Experimental Environment, Temperature Instrumentation, and Data Acquisition}
The MMRS was evaluated inside a 1~mm thick glass tube containing 35~g of distilled water, as shown in Fig.~\ref{fig:thermocouple_telemetry}, isolating the cooling hardware's performance from physiological variables like tissue heterogeneity and blood perfusion. While the 30~mm swimmer body operates on the millimeter scale to accommodate embedded thermocouple telemetry, it serves as a scaled physical proxy to evaluate the coupled electromagnetic-thermal physics inherent to untethered microrobotic workspaces.

Temperatures were monitored using non-magnetic, Type E thermocouples with a measurement range of $-270^{\circ}$C to $900^{\circ}$C. These sensors were chosen specifically to prevent magnetic field interference and eddy-current artifacts during induction heating~\cite{dempsey2001investigation}. The sensors were positioned at three key locations: the MMRS tip to measure primary heating, the surrounding water to characterize fluid temperature response, and the outer glass surface to characterize potential heating of neighboring tissue.

The thermocouples were interfaced with a National Instruments CompactRIO (cRIO) system equipped with an NI-9211 high-resolution analog input module featuring internal cold-junction compensation. This setup minimized electromagnetic noise from the induction heater while logging calibrated temperature data at a stable sampling frequency of 6~Hz.

\subsection{Thermal Performance Criteria}
Both configurations were subjected to induction heating for a duration of 100~s. Evaluation bounds were established to ensure sufficient therapeutic ablation at the target site while protecting surrounding biological structures. Given that the physiological baseline for human body temperature ranges from 36.1$^{\circ}$C to 37.2$^{\circ}$C~\cite{mackowiak1992critical}, clinical hyperthermia standards dictate that localized peripheral tissue can transiently tolerate exposure up to 43$^{\circ}$C before the onset of permanent cellular thermal damage~\cite{Dewey1994,Sapareto1984}. Consequently, to prevent collateral healthy tissue injury during short-duration procedures, the system must simultaneously satisfy three tiered operational constraints:
\begin{subequations}
\begin{align}
60~^{\circ}\mathrm{C} &< T_{\mathrm{MMRS}} < 95~^{\circ}\mathrm{C}, 
\label{eq:mmrs_temp_constraint}\\
T_{\mathrm{surface}} &< 43~^{\circ}\mathrm{C}, \label{eq:surface_temp_constraint}\\
T_{\mathrm{water}} &< 43~^{\circ}\mathrm{C}. \label{eq:water_temp_constraint}
\end{align}
\end{subequations}
where $T_{\mathrm{MMRS}}$ denotes the localized ablation temperature at the microrobot tip, while $T_{\mathrm{surface}}$ and $T_{\mathrm{water}}$ represent upper safety boundaries tracking the test tube surface and surrounding fluid environment, respectively, ensuring they remain strictly beneath the thermal damage ceiling.
\section{Experimental Results}
\begin{figure*}[!b]
  \centering
  \begin{minipage}{0.48\linewidth}
    \centering
    \includegraphics[width=1.0\linewidth]{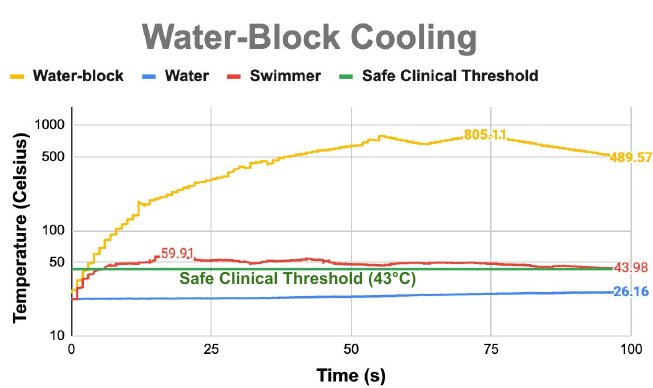}
    \caption{Transient thermal performance profiles over a 100 s heating cycle for the water-block configuration.}
    \label{fig:water_block_results}
  \end{minipage}\hfill
  \begin{minipage}{0.48\linewidth}
    \centering
    \includegraphics[width=1.0\linewidth]{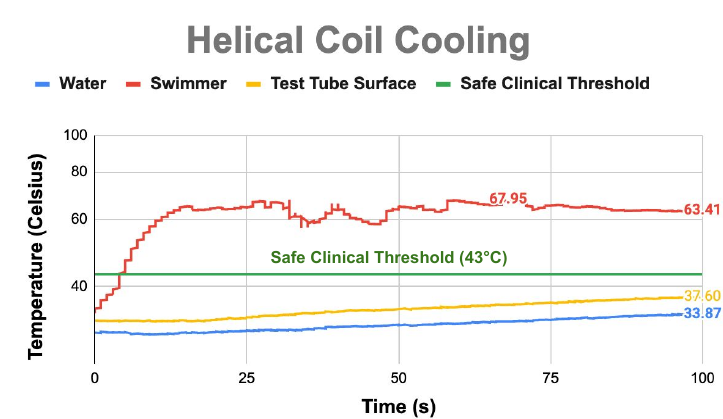}
    \caption{Transient thermal performance profiles over a 100 s heating cycle for the directly water-cooled coil configuration.}
    \label{fig:direct_water_results}
  \end{minipage}
  \vspace{-1.0em} 
\end{figure*}
\subsection{Water Block Configuration}
The evaluation of Configuration 1 revealed a severe operational failure mode where the conductive copper water block acted as an unintended inductive susceptor. As illustrated by the logarithmic profiles in Fig.~\ref{fig:water_block_results}, the block experienced rapid parasitic heating, exhibiting structural scorching by 15~s, visible thermal incandescence by 35~s, and a transient peak of 805$^{\circ}$C at 74~s. Because this extreme thermal signature vastly surpassed acceptable physiological boundaries and threatened to structurally melt the experimental apparatus, the induction power source was manually de-energized at the 74~s mark as an emergency safety intervention. Following this power shutdown, the block and swimmer temperatures declined to 489$^{\circ}$C and 44.0$^{\circ}$C at the 100~s terminal mark, respectively.

\subsection{Directly Water Cooled Configuration}
In contrast, the directly water-cooled coil configuration demonstrated exceptional thermal stability, successfully isolating the inductive heating to the target workspace. As illustrated by the logarithmic profiles in Fig.~\ref{fig:direct_water_results}, the MMRS tip temperature reached a transient peak of 68.0$^{\circ}$C at 68~s before stabilizing to 63.4$^{\circ}$C at the 100~s terminal mark. Simultaneously, active internal fluidics arrested parasitic thermal accumulation, maintaining the surrounding fluid at 33.9$^{\circ}$C while the external test tube surface stabilized at approximately 37.6$^{\circ}$C, well below critical damage thresholds.

This controlled thermal performance confirms that Configuration 2 achieves therapeutic ablation thresholds at the microrobot tip while rigorously restricting all peripheral boundaries below the critical 43$^{\circ}$C clinical damage threshold.

\section{Discussion}
The starkly contrasting thermal profiles underscore a fundamental design trade-off. In Configuration 1, the intermediate, contact-reliant copper block acted as an unintended inductive susceptor. This direct coupling to the high-flux coil generated intense electromagnetic losses that rapidly overwhelmed its internal fluidic cooling, initiating localized channel boiling and vapor-insulated thermal runaway that peaked at 805$^{\circ}$C. Rather than mitigating heat accumulation, the external block introduced a parasitic thermal node that fundamentally constrained the active workspace. Conversely, the MMRS tip's thermal trajectory remained strictly bounded below the defined threshold of the ambient water heat sink, verifying the insulation efficacy of direct internal coil fluidics.

While replacing the copper block with an insulating ceramic like aluminum nitride could bypass this failure mode, such design remains suboptimal; these bulky external structures constrain the active electromagnetic workspace and rely heavily on contact-reliant interface cooling. In contrast, the hollow internal fluidic loop of Configuration 2 extracts parasitic ohmic losses directly at the primary generation node, eliminating secondary induction while preserving a clear, unobstructed electromagnetic actuation workspace.

Configuration 2 addresses this failure mode by routing the coolant directly through the hollow conductor channel, effectively extracting heat at the precise location of parasitic ohmic dissipation. This direct internal extraction established a controlled, highly favorable temperature separation between the localized ablation zone and the outer boundary at the 100~s terminal mark, quantified by:
\begin{equation}
\Delta T_{\text{MMRS}-\text{surface}} 
= 63.4 - 37.6 
= 25.8^{\circ}\text{C}.
\label{eq:temperature_difference}
\end{equation}
This substantial temperature differential validates the efficacy of the direct cooling architecture, demonstrating its capacity to sustain the high localized hyperthermia required for therapeutic ablation while thermally isolating the surrounding environment.

\subsection{Limitations}
While these experiments validate the direct cooling concept, several limitations remain. 
\textbf{First}, the benchtop setup omits the heterogeneous thermal properties, microvascular perfusion, and complex geometry of living biological tissue. 
\textbf{Second}, data acquisition was constrained to discrete thermocouple point measurements; thus, the temperature differential in \eqref{eq:temperature_difference} reflects a separation between localized peak values rather than a full 3D temperature gradient field. Furthermore, due to the destructive nature of the thermal runaway in Configuration 1, iterative optimization trials were restricted to preserve system hardware. 
\textbf{Finally}, the multi-variable cross-coupling between coolant flow rate, input induction power, relative MMRS positioning, and spatial temperature distributions was not systematically mapped, which remains a key target to define the robust operating envelope of the automated platform.

\section{Conclusion}
This study demonstrated a coil-level liquid-cooling architecture capable of regulating the intense parasitic thermal behavior of an induction-heated Miniature Magnetic Robotic Swimmer (MMRS). While traditional copper water block configurations fail due to remote heat extraction, routing coolant directly through a hollow helical induction coil extracts parasitic losses at the source. This architecture enabled the MMRS to achieve a terminal therapeutic ablation temperature of 63.4$^{\circ}$C while restricting the surrounding surrogate fluids and test-tube surfaces below defined thresholds of 33.9$^{\circ}$C and 37.6$^{\circ}$C, respectively.

Although this 25.8$^{\circ}$C thermal separation indicates substantial insulation, it captures discrete point measurements rather than a continuous spatial gradient. Future work will focus on ex vivo tissue validation and transition the platform to autonomous, closed-loop PID regulation. To compensate for variable electromagnetic coupling, the controller will implement state-dependent feedback gains parameterized by the microrobot's 3D spatial location $\mathbf{r}(t)$, integrated with a parallel supervisory safety layer to instantly cut power if operational boundaries are breached.

Ultimately, coil-level heat extraction provides a scalable, stable mechanical foundation for localized, high-temperature thermal ablation using remotely actuated magnetic microrobots.
\section*{Acknowledgment}
The authors thank the National Science Foundation (NSF) for supporting this research through the Research Experience for Mentoring (REM) program. We also thank the Department of Electrical and Computer Engineering at the University of Houston for facility and hardware support.
\begin{small}
\bibliographystyle{IEEEtran}
\bibliography{ref}
\end{small}
\end{document}